# Analytical solutions for optical forces between two dielectric planar waveguides immersed in dielectric fluid media

**Janderson Rocha Rodrigues[1,2,*] and Vilson Rosa Almeida[1,2]**

[1]*Instituto Tecnológico de Aeronáutica, 50 Praça Marechal Eduardo Gomes, Vila das Acacias, São José dos Campos - SP. 12228-900, Brazil*
[2]*Instituto de Estudos Avançados, 01 Trevo Cel Av José Alberto Albano do Amarante, Putim, São José dos Campos - SP. 12228-001, Brazil*
[*]*jrr@ita.br*

**Abstract**

We investigate optical (transverse gradient) forces between two high-index dielectric planar waveguides immersed in low-index dielectric fluid media. Complimentary to previous studies, we extend optical forces calculations, in order to take into account a non-vacuum (and non-air) background medium, by using the Minkowski stress tensor formulation; we derived a very simple set of equations in terms of the effective refractive indexes of the waveguide eigenmodes. We also used a normalized version of the dispersion relation method to calculate the optical forces, in order to validate our results for different dielectric fluid media. Excellent agreement between the two methods was obtained for all analyzed cases. We show that, due to slot-waveguide effect, the TM modes are more sensitive to changes in the fluid refractive index than the TE ones. Furthermore, the repulsive optical force of the antisymmetric $TM_1$ mode becomes stronger for higher refractive indexes, whereas the attractive force of the symmetric $TM_0$ mode becomes weaker. The methodology and results presented in this work provide a rigorous analysis of nano-optomechanical devices actuated by optical forces in a broader range of materials and applications.

## 1. Introduction

Optical (transverse gradient) forces between two parallel dielectric rectangular waveguides were theoretically investigated by Povinelli *et al.* [1]. The optical force originates from dipole moments induced in the dielectric waveguides by the light intensity of the guided electromagnetic fields. Such forces may be either attractive or repulsive, depending on the relative phase of the dipoles, dictated by the optical properties and the geometrical parameters of the waveguides. The symmetric modes lead to attractive (bonding) optical forces, pulling the waveguides together, whereas the antisymmetric modes lead to repulsive (antibonding) forces, pushing the waveguides apart. These optical forces have been experimentally demonstrated in several nanophotonics dielectric structures, such as: between a silicon nitride micro-disk and a silica waveguide [2], between a silicon waveguide and a silicon oxide substrate [3], between two silicon waveguides [4,5], in resonant silicon nitride structures [6], among many others [7,8].

The assessment of the optical forces acting between two non-absorbing (lossless) homogeneous and isotropic dielectric structures may be obtained from the device's dispersion relation (DR) as a function of their separation [1]. Rigorously, these optical forces are calculated using the Maxwell stress tensor (ST) formalism [9]. This method allows the calculation of the total electromagnetic force acting on an object by integrating the Maxwell

ST, which is a function of the electromagnetic field spatial distributions, around an arbitrary closed surface surrounding the object. The Maxwell ST formalism requires that the inclosing-object surface lies completely in vacuum or air [9]. This method was used explicitly to evaluate optical forces between two dielectric optomechanical devices in: reflectors [10], mirrors [11], planar waveguides [12], waveguide & substrate [13], and waveguide & ring resonator [14], among others. In all these cases, the authors have assumed air as the dielectric fluid medium between the two components, which is an imposed condition for using of the Maxwell ST [9].

On the other hand, these nano-optomechanical devices may be subjected to innumerous practical applications where the dielectric surround medium is not necessarily vacuum (or air) as, for example, water or ethanol in biological and/or chemical sensing mechanism, or manipulation of dielectric nanoparticles in optofluidic actuation mechanisms [15,16]. Therefore, extending the approaches adopted in previous works, here we extend the optical forces calculus for a general surrounding dielectric fluid medium, by applying the Minkowski ST formulation in the case of two dielectric planar waveguides. We also calculate the optical force using a normalized version of the DR method. Finally, we apply both methods for different fluid-surround media and compare the results.

## 2. Two dielectric planar waveguides

We consider the geometry shown schematically in Fig. 1. The dielectric structure is formed by two planar waveguides of high refractive index material ($n_H$), separated by a gap $g = 2a$, and surrounded by a low refractive index material ($n_L$). The waveguides width is $w = b - a$ (assuming $b > a > 0$) and thickness $t$ and length $L$, and are symmetrical in both the $x$- and $y$-direction.

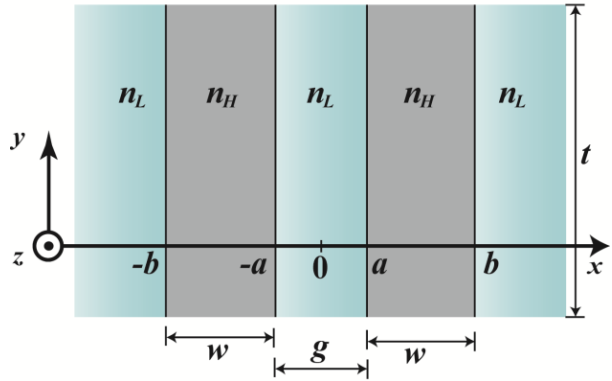

Fig. 1. Schematic view of the two high-index ($n_H$) dielectric planar waveguides surrounded by a low-index ($n_L$) dielectric fluid.

Here, we assume that the waveguides are semi-infinite in extent in the $y$-direction ($t \gg w$), translational invariance of the structure in the propagation direction $z$, and it is excited by a harmonic source with a fixed angular frequency $\omega = \omega_0$. Maxwell's equations, under these assumptions, lead to two independent sets of solutions for the field polarization: Transverse Electric (TE) or Transverse Magnetic (TM), depending on the orientation of the main component of the electric field. In the TE case, the electric field is polarized along the $y$-axis $(0,0,E_y,0,H_x,H_z)$ and in the TM one along the $x$-axis $(E_x,0,E_z,0,H_y,0)$.

The $E_y(x)$ component of TE polarization is given by solving the following transverse wave equation,

$$\frac{\partial^2 E_y(x)}{\partial x^2} + (k_0^2 n_{sf}^2 - \beta_{TE}^2) E_y(x) = 0 \tag{1}$$

where $\beta_{TE}$ is the eigenmode propagation constant along the z-direction (or longitudinal wavevector), $k_0$ is the vacuum (or free space) wavevector magnitude, which is related to the angular frequency $\omega_0$ and the speed of light in vacuum $c_0$ through $k_0 = \omega_0/c_0 = \omega_0\sqrt{\mu_0\varepsilon_0}$, where $\varepsilon_0$ and $\mu_0$ are the vacuum electric permittivity and the magnetic permeability, respectively. The refractive index $n_{sf}$ is a step-index function, which is equal to $n_L$, for $|x| < a$ or $|x| > b$, and to $n_H$ for $a \leq |x| \leq b$.

In the TE polarization, the spatial-distribution components of the fields are related to each other through:

$$[E_y(x)\ H_x(x)\ H_z(x)]^T = \left[1\ -\frac{\beta_{TE}}{\omega_0\mu_0}\ \frac{i}{\omega_0\mu_0}\frac{\partial}{\partial x}\right]^T E_y(x) \tag{2}$$

where $[\cdot]^T$ denotes matrix transpose.

Similarly, the $H_y(x)$ component of TM polarization is given by solving the following equation,

$$\frac{\partial^2 H_y(x)}{\partial x^2} + (k_o^2 n_{sf}^2 - \beta_{TM}^2) H_y(x) = 0 \tag{3}$$

where the spatial-distribution components are given by:

$$[E_x(x)\ E_z(x)\ H_y(x)]^T = \left[\frac{1}{n_{sf}^2}\frac{\beta_{TM}}{\omega_0\varepsilon_0}\ \frac{1}{n_{sf}^2}\frac{1}{\omega_0\varepsilon_0}\frac{\partial}{\partial x}\ 1\right]^T H_y(x) \tag{4}$$

The solutions of the Equations (1) and (3) are obtained by applying the following boundary conditions: the continuity of the tangential electric and magnetic fields (and their derivatives) at the dielectric interfaces ($x = |a|$ and $x = |b|$) and energy conservation requirements. Then, the $E_y(x)$ and $H_y(x)$ components for the TE and TM polarizations, respectively, can be expressed as:

$$E_y(x) = \frac{A_{TE}}{2}\begin{cases} \left(e^{(\gamma_L x)} \pm e^{-(\gamma_L x)}\right), & |x| \leq a \\ \left(e^{(\gamma_L a)} \pm e^{-(\gamma_L a)}\right)\cos[\kappa_H(|x|-a)] + \\ +\frac{\gamma_L}{\kappa_H}\left(e^{(\gamma_L a)} \mp e^{-(\gamma_L a)}\right)\sin[\kappa_H(|x|-a)], & a \leq |x| \leq b \\ \left\{\left(e^{(\gamma_L a)} \pm e^{-(\gamma_L a)}\right)\cos[\kappa_H w] + \right. \\ \left. +\frac{\gamma_L}{\kappa_H}\left(e^{(\gamma_L a)} \mp e^{-(\gamma_L a)}\right)\sin[\kappa_H w]\right\}e^{-\gamma_L(x-b)}, & |x| \geq b \end{cases} \tag{5}$$

and

$$H_y(x) = \frac{A_{TM}}{2} \begin{cases} \left(e^{(\gamma_L x)} \pm e^{-(\gamma_L x)}\right), & |x| \leq a \\ \left(e^{(\gamma_L a)} \pm e^{-(\gamma_L a)}\right)\cos[\kappa_H(|x|-a)] + \\ +\dfrac{n_H^2}{\kappa_H}\dfrac{\gamma_L}{n_L^2}\left(e^{(\gamma_L a)} \mp e^{-(\gamma_L a)}\right)\sin[\kappa_H(|x|-a)], a \leq |x| \leq b \\ \left\{\left(e^{(\gamma_L a)} \pm e^{-(\gamma_L a)}\right)\cos[\kappa_H w] + \right. \\ \left. +\dfrac{n_H^2}{\kappa_H}\dfrac{\gamma_L}{n_L^2}\left(e^{(\gamma_L a)} \mp e^{-(\gamma_L a)}\right)\sin[\kappa_H w]\right\} e^{-\gamma_L(|x|-b)}, |x| \geq b \end{cases} \quad (6)$$

where $A_{TE}$ and $A_{TM}$ are the field amplitude at the position $x = 0$. Due to the structure symmetry in relation to the $x$-plane, the electromagnetic field distributions for each mode can be classified into symmetric/even or antisymmetric/odd. In order to represent both solutions, we have used the mathematical relations $2\cosh\theta = e^\theta + e^{-\theta}$ and $2\sinh\theta = e^\theta - e^{-\theta}$, and the signs "$\pm$" and "$\mp$" to represent this classification, where the top sign represents the symmetric modes and the bottom sign represents the antisymmetric ones. We use this notation convention hereinafter in this work.

The transcendental characteristic equations for both the TE and TM polarizations are given, respectively, by.

$$\tanh(\gamma_L a) = \left\{\frac{\kappa_H^2 \tan[\kappa_H w] - \kappa_H \gamma_L}{\gamma_L^2 \tan[\kappa_H w] + \kappa_H \gamma_L}\right\}^{\pm 1} \quad (7)$$

and

$$\tanh(\gamma_L a) = \left\{\frac{\dfrac{\kappa_H^2}{n_H^4}\tan[\kappa_H w] - \dfrac{\kappa_H}{n_H^2}\dfrac{\gamma_L}{n_L^2}}{\dfrac{\gamma_L^2}{n_L^4}\tan[\kappa_H w] + \dfrac{\kappa_H}{n_H^2}\dfrac{\gamma_L}{n_L^2}}\right\}^{\pm 1} \quad (8)$$

where $\kappa_H$ and $\gamma_L$ are the transversal wavenumber and the field decay coefficient, respectively given by:

$$\kappa_H = \sqrt{k_0^2 n_H^2 - \beta^2} \quad (9)$$

and

$$\gamma_L = \sqrt{\beta^2 - k_0^2 n_L^2} \quad (10)$$

The numerical solutions of Eqs. (1) and (2) yield the eigenvalues of $\beta$ that correspond to allowed guided modes in the waveguide for each polarization. For these modes, the universal condition $n_L k_0 < \beta < n_H k_0$ or $n_L < n_{eff} < n_H$ must be satisfied, assuming $n_L \leq n_H$, where $n_{eff}$ is effective refractive index, defined by $n_{eff} \equiv \beta/k_0$. Here, we restrict our analysis only to guided modes, disregarding the radiation modes ($n_{eff} < n_L$).

The fields' amplitudes ($A_{TE}$ and $A_{TM}$) are related to the optical power carried by each mode in the waveguide. This power is calculated by integrating the real part of time-averaged

$z$-component of the Poynting vector over the waveguide cross-sectional area. The optical power per unit length in the $y$-direction for the TE and TM polarization modes are calculated respectively by:

$$\frac{P_{zTE}}{t} = \frac{1}{2}\frac{\beta_{TE}}{\omega_0\mu_0}\int_{-\infty}^{\infty}|E_y(x)|^2 dx \tag{11}$$

and

$$\frac{P_{zTM}}{t} = \frac{1}{2}\frac{\beta_{TM}}{\omega_0\varepsilon_0}\int_{-\infty}^{\infty}\frac{1}{n_{sf}^2}|H_y(x)|^2 dx \tag{12}$$

By substituting Eqs. (5) and (6) into Eqs. (11) and (12), it is possible to express analytically the linear density of optical power $P_z/t$ for each polarization, as a function of the field amplitudes' coefficients $A_{TE}$ and $A_{TM}$, respectively:

$$\frac{P_{zTE}}{t} = \frac{2n_{effTE}}{c_0\mu_0}|A_{TE}|^2\left\{\left[\frac{\sinh(2\gamma_L a)}{4\gamma_L}\pm\frac{a}{2}\right]+\right.$$

$$+\left[\frac{\sin(2\kappa_H w)}{4\kappa_H}\left(\frac{\cosh(2\gamma_L a)}{2}\pm\frac{1}{2}\right)+\frac{w}{2}\left(\frac{\cosh(2\gamma_L a)}{2}\pm\frac{1}{2}\right)+\frac{\gamma_L}{2\kappa_H^2}\sinh(2\gamma_L a)\sin^2(\kappa_H w)+\right.$$

$$-\frac{\gamma_S^2}{4\kappa_H^3}\sin(2\kappa_H w)\left(\frac{\cosh(2\gamma_L a)}{2}\mp\frac{1}{2}\right)+\frac{\gamma_S^2 w}{2\kappa_H^2}\left(\frac{\cosh(2\gamma_L a)}{2}\mp\frac{1}{2}\right)\right]+$$

$$\left.+\frac{1}{2\gamma_L}\left[\cos(\kappa_H w)\left(\frac{\cosh(2\gamma_L a)}{2}\pm\frac{1}{2}\right)+\frac{\gamma_L}{\kappa_H}\sin(\kappa_H w)\left(\frac{\cosh(2\gamma_L a)}{2}\mp\frac{1}{2}\right)\right]^2\right\} \tag{13}$$

and

$$\frac{P_{zTM}}{t} = \frac{2n_{effTM}}{c_0\varepsilon_0}|A_{TM}|^2\left\{\frac{1}{n_L^2}\left[\frac{\sinh(2\gamma_S a)}{4\gamma_S}\pm\frac{a}{2}\right]+\right.$$

$$+\frac{1}{n_H^2}\left[\frac{\sin(2\kappa_H w)}{4\kappa_H}\left(\frac{\cosh(2\gamma_L a)}{2}\pm\frac{1}{2}\right)+\frac{w}{2}\left(\frac{\cosh(2\gamma_L a)}{2}\pm\frac{1}{2}\right)+\right.$$

$$+\frac{n_H^2}{2n_L^2}\frac{\gamma_S}{\kappa_H^2}\sinh(2\gamma_S a)\sin^2(\kappa_H w)+$$

$$-\frac{n_H^4}{n_L^4}\frac{\gamma_L^2}{4\kappa_H^3}\sin(2\kappa_H w)\left(\frac{\cosh(2\gamma_L a)}{2}\mp\frac{1}{2}\right)+\frac{n_H^4}{n_L^4}\frac{\gamma_L^2 w}{2\kappa_H^2}\left(\frac{\cosh(2\gamma_L a)}{2}\mp\frac{1}{2}\right)\right]+$$

$$\left.+\frac{1}{2\gamma_L n_L^2}\left[\cos(\kappa_H w)\left(\frac{\cosh(2\gamma_L a)}{2}\pm\frac{1}{2}\right)+\frac{n_H^2}{n_L^2}\frac{\gamma_L}{\kappa_H}\sin(\kappa_H w)\left(\frac{\cosh(2\gamma_L a)}{2}\mp\frac{1}{2}\right)\right]^2\right\} \tag{14}$$

## 3. Optical forces between two planar waveguides

By using linear-momentum conservation arguments, it is possible to show that, for time-harmonic fields, the time-averaged of the total electromagnetic force on a dielectric object is [17],

$$F_i(\vec{r}) = -\oint_A \langle T_{ij}(\vec{r})\rangle dA_j = -\frac{1}{2}Re\left\{\oint_A T_{ij}(\vec{r})dA_j\right\} \quad (15)$$

where $i$ and $j$ represent the Cartesian coordinates $\{x,y,z\}$, $Re\{.\}$ is the real operator, and $\vec{A}$ is an outward normal pointing vector on the surface enclosing the object volume. The $\overleftrightarrow{T}$ is the force per unit area (stress or pressure) tensor acting on the surface. More precisely, $T_{ij}$ is the force on the $i$-th direction acting on an element of area oriented in the $j$-th direction [9].

The optical forces are usually calculated using Maxwell ST [10-14]; however, the Maxwell ST requires that the inclosing-object surface $A$ lies completely in vacuum (or in air) [9,17]. Here, we have applied the Minkowski ST formulation in order to calculate for the optical forces between dielectric objects embedded in dielectric fluid, assuming mechanical and thermal equilibrium. The same procedure has been used to calculate optical forces in dielectric devices immersed in dielectric media subjected to transverse (normal or oblique) light incidence [18,19] and in non-movable photonic structures [15,16]; in contrast, here we are dealing with guided waves in optomechanical devices.

The Minkowski ST is given by [17]:

$$T_{ij}(\vec{r}) = \left[E_i(\vec{r})D_j^*(\vec{r}) + H_i(\vec{r})B_j^*(\vec{r}) - \frac{1}{2}\delta_{ij}\sum_k \left(E_k(\vec{r})D_k^*(\vec{r}) + H_k(\vec{r})B_k^*(\vec{r})\right)\right] \quad (16)$$

where the fields are represented as complex vector components, the symbol "*" denotes their complex conjugate, $\delta_{ij}$ is the Kronecker delta, and $\vec{D}$ and $\vec{B}$ are the electric displacement and the magnetic flux density, respectively. For a homogeneous isotropic and linear dielectric medium, $\vec{D} = \varepsilon\vec{E}$ and $\vec{B} = \mu\vec{H}$, where $\varepsilon$ and $\mu$ are the material electric permittivity and magnetic permeability, respectively. The Minkowski ST reduces to the Maxwell ST in vacuum (or in air): $\varepsilon = \varepsilon_0$ and $\mu = \mu_0$. Consider a non-magnetic medium, $\mu = \mu_0$, we have $\varepsilon = n_L^2$, therefore the component of force in the $x$-direction acting on an element of area oriented in the same direction is given by,

$$\langle T_{xx}(x)\rangle = -\frac{\varepsilon_0}{2}\left[n_L^2\left(|E_y(x)|^2 + |E_z(x)|^2 - |E_x(x)|^2\right) + \right.$$

$$\left. +c^2\mu_0^2\left(|H_y(x)|^2 + |H_z(x)|^2 - |H_x(x)|^2\right)\right] \quad (17)$$

which can be easily separated between both polarization,

$$\langle T_{xx,TE}(x)\rangle = -\frac{\varepsilon_0}{4}\left[n_L^2|E_y(x)|^2 + c^2\mu_0^2(|H_z(x)|^2 - |H_x(x)|^2)\right] \quad (18)$$

and

$$\langle T_{xx,TM}(x)\rangle = -\frac{\varepsilon_0}{4}\left[n_L^2|E_z(x)|^2 - n_L^2|E_x(x)|^2 + c^2\mu_0^2\left(|H_y(x)|^2\right)\right] \quad (19)$$

The optical forces, per unit area, are evaluated by means of the Minkowski ST in the gap region between the two planar waveguides, considering that the force is always oriented with respect to the surface normal direction and that $g = 2a$, given by

$$\frac{F_{optx}(g)}{t\,L} = -\langle T_{xx}(x)\rangle\Big|_{x=-a^+} - \langle T_{xx}(x)\rangle\Big|_{x=+a^-} = -2\langle T_{xx}(x)\rangle\Big|_{x=g/2} \tag{20}$$

Substituting the field distributions in Eqs. (18) and (19) and replacing them in Eq. (20), the optical forces per unit area for the TE and TM polarizations are given, respectively, by

$$\frac{F_{optTE}(g)}{t\,L} = \pm\frac{\varepsilon_0}{2}|A_{TE}|^2\,n_L^{\,2}\left(1 - \frac{n_{effTE}(g)^2}{n_L^{\,2}}\right) \tag{21}$$

and

$$\frac{F_{optTM}(g)}{t\,L} = \pm\frac{\mu_0}{2}|A_{TM}|^2\left(1 - \frac{n_{effTM}(g)^2}{n_L^{\,2}}\right) \tag{22}$$

From Eqs. (21) and (22) one can see that, since for guided modes $n_{eff} > n_L$, the optical forces will be always attractive (negative) for the symmetric mode and always repulsive (positive) for antisymmetric ones, for both polarizations, regardless of the value of $n_L$. Furthermore, when the two waveguides are far away from each other, they became degenerate, acting as two isolated waveguides; as they come closer ($g \ll 1/\gamma_L$) they become highly coupled, forming a slab-based slot-waveguide; for the TM polarization, there is a further confinement and enhancement of the light in the gap region [21], which increases the sensitivity with $n_L$. Setting $n_L = 1.0$ in Eqs. (21) and (22), turn them into the same results presented in the equations 11 and 12 from reference [12], and in the equations 18 and 19 from reference [20], which were obtained using the Maxwell ST formulation. Therefore, the results presented in the present work generalize the optical forces calculus for any value of $n_L$.

On the other hand, by using Newton's second law, the work-energy principle, and the first law of thermodynamics, considering an adiabatic process, one can show that the time-averaged the optical force between two dielectric waveguides can be alternatively calculated from the dispersion relation (DR) of these waveguides [1]. This force may be expressed through the derivative of the mode effective refractive index, $n_{eff}$, with respect to the gap distance, $g$ [13]:

$$F_{opt}(g) = -\frac{1}{c_0}\frac{n_g(g)}{n_{eff}(g)}\frac{\partial n_{eff}(g)}{\partial g}\bigg|_\beta P_z\,L \tag{23}$$

where $n_g$ is the group refractive index. The derivative of the eigenmode effective refractive index with respect to the gap is calculated for a fixed longitudinal propagation constant $\beta$, i.e. for a unique guided optical longitudinal wavenumber. The relationship between the group refractive index and the effective refractive index in terms of the angular frequency is given by,

$$n_g(\omega) = n_{eff}(\omega) - \omega\frac{dn_{eff}(\omega)}{d\omega} \tag{24}$$

By using Eq. (23), the relation $\omega_0 = c_0\beta/n_{eff}$, and the triple product rule, it is possible to show that:

$$\left.\frac{\partial n_{eff}(g)}{\partial g}\right|_\beta = -\frac{n_{eff}(g)}{n_g(g)}\left.\frac{\partial n_{eff}(g)}{\partial g}\right|_{\omega_0} \qquad (25)$$

Replacing Eq. 25 into Eq. 23, the optical force reduces to

$$F_{opt}(g) = \frac{1}{c_0}\left.\frac{\partial n_{eff}(g)}{\partial g}\right|_{\omega_0} P_z L \qquad (26)$$

where the derivative of the mode effective refractive index with respect to the gap is now performed for a fixed angular frequency $\omega_0$, i.e. for a fixed wavelength $\lambda_0$, where $\omega_0 = 2\pi c_0/\lambda_0$. The relation showed in Eq. (26) can also be obtained through the analysis of the optical phase and amplitude responses for the optomechanical system, due to changes in the its mechanical parameters, by using RTOF (*Response Theory of Optical Forces*) [22]. Despite of the fact that one method is based on a closed-system [1] and the other on an open-system [22], we obtain exact equivalence between both methods [22].

The relationship described in the Eq. (23) or (26) was firstly established for rectangular waveguides [1]; however, we assume that the DR method is general, since it is derived from first principles, and can be applied to any structure, as well as for the planar case. Furthermore, as the effective refractive index for a given guided mode depends on the refractive index of the surrounding/cladding material ($n_L$), equation (26) must also take into account the non-vacuum or non-air case. In order to transform the mode optical power into power linear density, i.e. power per unit length in *y*-direction, we normalize it by the waveguides thickness $t$. Therefore, the optical forces per unit area using DR method for the TE and TM polarizations are given, respectively, by:

$$\frac{F_{optTE}(g)}{t\,L} = \frac{1}{c_0}\left.\frac{\partial n_{effTE}(g)}{\partial g}\right|_{\omega_0} \frac{P_{zTE}}{t} \qquad (27)$$

and

$$\frac{F_{optTM}(g)}{t\,L} = \frac{1}{c_0}\left.\frac{\partial n_{effTM}(g)}{\partial g}\right|_{\omega_0} \frac{P_{zTM}}{t} \qquad (28)$$

where the differentiation between the symmetric and the antisymmetric modes is given directly by the sign of their respective derivatives, where $\partial n_{eff}/\partial g < 0$ for symmetric modes and $\partial n_{eff}/\partial g > 0$ for antisymmetric one.

We use the same parameters used in [12]: two silicon planar waveguides ( $n_H = 3.5$ ), at a wavelength of $\lambda_0 = 1550$ nm, and width of $w = 310$ nm. Figure 2 shows the effective refractive indexes for the symmetric and antisymmetric modes for both polarizations, TE in Fig. 2(a) and TM in Fig. 2(b), and for three dielectric fluid media: air ( $n_L = 1.0$ ), water ( $n_L = 1.3$ ), and Cassia oil ( $n_L = 1.6$ ). In general, the effective refractive indexes increase for higher medium indexes. However, the effective indexes of the TM modes are more sensitive than the TE ones, i.e. the latter present larger variations than the former ones, for the same changes in medium index. The largest mode sensitivity occurs for the symmetric

(fundamental) TM mode, due to the field enhancement effect in slot-waveguide. At larger distances, the symmetric and antisymmetric effective indexes asymptotically converge to the refractive index of a waveguide comprised of a half-side of the slot waveguide.

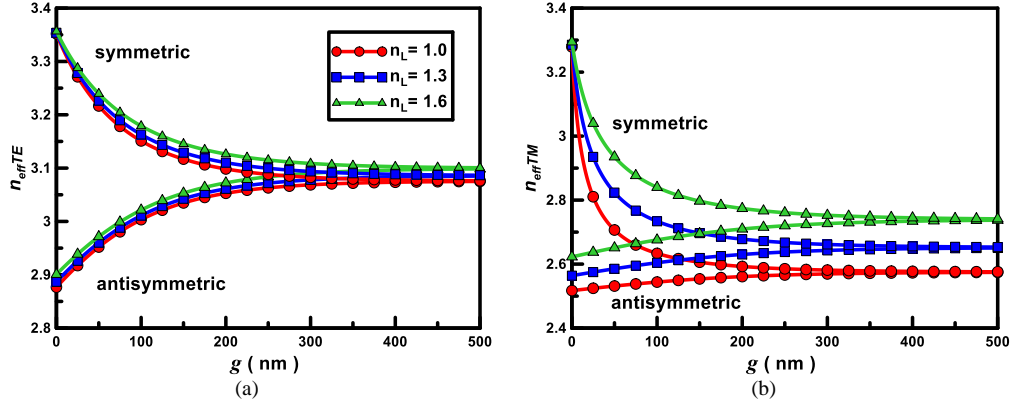

Fig. 2. Effective refractive indexes for the symmetric and the antisymmetric modes. (a) TE polarization. (b) TM polarization.

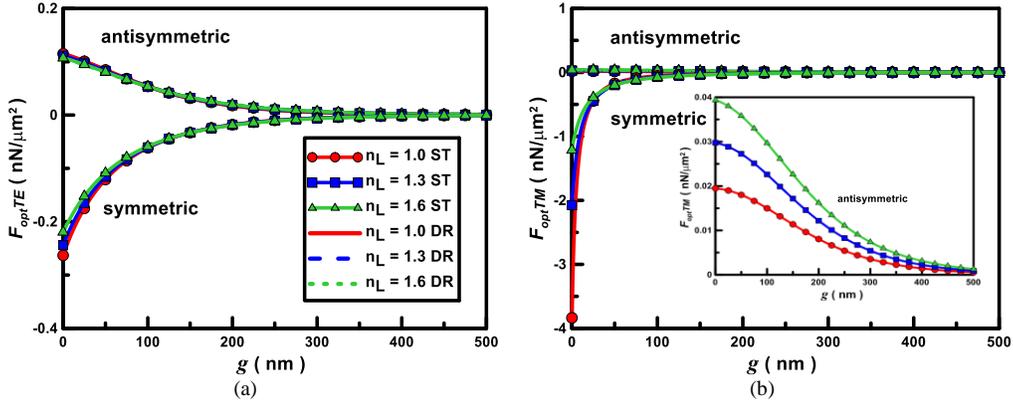

Fig. 3. Comparison between the optical forces obtained by Minkowski stress tensor (ST) (solid lines) and dispersion relation (DR) (dashed lines) for a power linear density of $P_z/t = 20 \ mW/\mu m$ for three dielectric fluid medium. (a) TE polarization. (b) TM polarization. Inset: antisymmetric TM mode.

Figure 3 compares the optical forces between the two planar waveguides obtained by both methods, Minkowski stress tensor (ST) and dispersion relation (DR), for a power linear density of $P_z/t = 20 mW/\mu m$ and for the three media. The optical force for the TE polarization is shown in Fig. 3(a), as well as the TM one in Fig. 3(b). Excellent agreement is obtained between the two methods for all analyzed cases. Furthermore, the optical forces decrease as $n_L$ increases. Due to the slot-waveguide effect, the largest (attractive) optical force is reached by the symmetric (fundamental) TM mode in air, and it becomes weaker as the medium index increases, since the contrast between the transverse electric field ($E_x(x)$) amplitude at the interface is proportional to the ratio $n_H^2/n_L^2$ [21]. Besides that, the term between parentheses in Eq. (21) reduces when the medium index increases.

On the other hand, as shown in the inset of Fig. 3(b), the repulsive optical force of the antisymmetric TM mode increases as the medium index increases. This unique behavior can be understood by the spatial distributions of the antisymmetric modes fields ($E_x(x)$ and

$H_y(x)$) and their hyperbolic sine shape in the gap region. By analyzing the antisymmetric version of Eq. (14) it is possible to notice that all terms containing hyperbolic sine goes to zero as $a$ goes to zero, in the center of the gap ($x = 0$), therefore, the remaining terms (last two lines) show that the field amplitude $|A_{TM}|^2$ of the antisymmetric TM mode is almost proportional to $n_L^4$. However, even with the gain in the field amplitude, when the medium index increases the term in parentheses in Eq. (22) decrease, limiting the gain in repulsive force. It is important to highlight that this effect of increasing the repulsive optical force for the antisymmetric TM mode can be explained only by the equations obtained by the ST method. Despite that, this effect may be used for tailoring the Casimir and van der Waals forces in these devices with less optical powers [23-24], since the dispersive forces also reduce in a non-vacuum or non-air fluid medium, in addition to the other applications mentioned before.

## 4. Conclusion

In this work, we analyzed optical forces between two dielectric planar waveguides immersed in dielectric fluid media by using the Minkowski stress tensor. We derived a very simple set of equations as a function of the modes effective refractive indexes, which recover the literature previous results in vacuum or in air condition. We also compared our results with the ones obtained through the dispersion relation method for three dielectric fluid media. Excellent agreement was obtained between the two methods. The methodology and results presented will allow one to analyze nano-optomechanical devices actuating by optical forces in a broad range of applications.

### Acknowledgments

This work was supported in part by the Coordenação de Aperfeiçoamento de Pessoal de Nível Superior (CAPES) through doctoral scholarship for J. R. Rodrigues and visiting professor sponsorship for V. R. Almeida, and in part by the Conselho Nacional de Desenvolvimento Científico e Tecnológico (CNPq) under Grant 312029/2013-6 and Grant 483116/2011-4.